\documentstyle[12pt,aaspp4]{article}

\def\etal{{\it et~al.\ }}

\tightenlines

\begin{document}

\title{Variable Extended Objects in $SA$ 57}

\author{Matthew A. Bershady\altaffilmark{1,2}} 

\affil{Department of Astronomy \& Astrophysics, Pennsylvania State
University, 525 Davey Lab, University Park, PA 16802 \\ and \\
Department of Astronomy, University of Wisconsin, 475 N. Charter Street,
Madison, WI 53706 (mab@astro.wisc.edu)}

\author{Dario Trevese\altaffilmark{1}}

\affil{Istituto Astronomico, Universita degli Studi ``La Sapienza,'' via
G.M. Lancisi 29, 00161, Rome, Italy (trevese@astrom.astro.it)}

\author{Richard G. Kron\altaffilmark{1}}

\affil{Fermi National Accelerator Laboratory, MS 127, Box 500,
Batavia, IL 60510 (rich@oddjob.uchicago.edu)}

\altaffiltext{1}{Visiting Astronomer, Kitt Peak National Observatory, 
National Optical Astronomy Observatories, which is operated by the
Association of Universities for Research in Astronomy, Inc. (AURA)
under cooperative agreement with the National Science Foundation}

\altaffiltext{2}{NOAO WIYN Queue Investigator}

\begin{abstract}

We have isolated a sample of 14 candidate variable objects with
extended image structure to $B_J = 22.5$ in 0.284 deg$^2$ of Selected
Area 57. The majority of candidates are blue ($U-B<0$) and relatively
compact. At fainter magnitudes, there is a steep rise in the number of
variable objects. These objects are also compact and blue, and some of
them are likely to be truly stellar. Twelve of the $B_J \leq 22.5$
candidates have been observed spectroscopically over limited ranges of
wavelength and a variety of resulting signal-to-noise. Three of the
four brightest have broad emission lines characteristic of Seyfert 1
galaxies. The fourth has a highly variable spectrum consistent with
Seyfert-like activity.  A fifth candidate has emission line ratios
nominally consistent with a Seyfert 2 galaxy. In most cases where we
have not been able to confirm a Seyfert spectroscopic type, the
spectra are of insufficient quality or coverage to rule out such a
classification. Based on spectroscopic and photometric redshift
information, the majority of candidates have luminosities less than
10\% of the nominal demarkation between QSOs and AGN (M$_B = -23$,
H$_0=50$ km s$^{-1}$ Mpc$^{-1}$, q$_0=0.5$). The surface density of
{\it confirmed} M$_B > -23$ AGN to $B_J = 22$, including stellar
sources, is $\sim$ 40 deg$^{-2}$, in good agreement with other surveys
at this depth. The confirmed AGN in extended sources make up 36\% of
this population. Thus, the application of a variability criterion to
images with extended structure enhances the completeness of the census
of active nuclei. If the majority of our candidates are bona fide AGN,
the surface density could be as high as 82 deg$^{-2}$ for M$_B > -23$,
and 162 deg$^{-2}$ for all luminosities to $B_J = 22$, with extended
sources contributing up to 33\% of the total.

\end{abstract}


\clearpage

\section{Background}

We develop techniques for the identification of low-luminosity active
galactic nuclei based on their variability.  This effort is a direct
extension of the survey for variability in images with stellar
structure by Trevese \etal (1994; T94).  Here, we consider images
that are detected to be extended, which would be expected to
correspond to Seyfert 1 galaxies.

In a series of papers (Kron \& Chiu 1981, Koo, Kron, \& Cudworth 1986
[KKC], Koo \& Kron 1988 [KK88], Trevese \etal 1989 [T89], Majewski
\etal 1991, and Trevese \etal 1994 [T94]), we have elaborated a number
of search techniques for quasars in Selected Area 57 to faint limits
($B = 22.5$), including colors, lack of proper motion, and
variability.  These various techniques yield somewhat different
samples, which then serve to calibrate the incompleteness of any
single technique.  However, all of these studies have relied on a
parent sample that is restricted to images with structures consistent
with star images.  One of our results is that the sample of apparently
faint quasars is rich in objects that are absolutely faint: according
to T94, the sample of 35 quasars has a median luminosity of M$_B$ =
-23.9 (H$_0$ = 50 km s$^{-1}$ Mpc$^{-1}$, q$_0$ = 0.5, $\alpha$ = -1),
and three quasars have luminosities fainter than M$_B$ = -22.0.  Five
have redshifts smaller than 0.8.  Traditionally Seyfert 1 galaxies and
quasars are separated at M$_B$ = -23 (a distinction we have previously
ignored by calling all of the objects in our lists with stellar images
and broad lines ``quasars'').  Clearly there is a likelihood that we
have missed some active nuclei in $SA$ 57 because we have only
searched for objects that have stellar image structure.  The point of
the present investigation is to address this issue, with the goal of
extending the completeness of our survey to still fainter absolute
magnitudes.

Another result of T94 was that essentially all of the known quasars
were detected to be variable, and moreover we argued that the
characteristic fractional variability amplitude was generally somewhat
higher for active nuclei of lower luminosity.  These findings suggest
that Seyfert galaxies could be discovered by the property of being
variable. Variability is well known to be a common property of
Seyfert 1 galaxies, but samples of Seyfert galaxies have been
constructed using some other defining criterion, such as spectroscopic
characteristics (Cheng \etal 1985; Huchra \& Burg 1992) or X-ray flux
(Maccacaro \etal 1991).  Hawkins (1983) has consistently advocated
the utility of variability as a technique for discovery of quasars.
His study, like the present one, was not restricted to images with
stellar structure, and he, too, detected some extended images to be
variable.

Aside from enhancing the completeness of our catalogue of active
nuclei in $SA$ 57, there are some more specific motivations to
investigate the space density of Seyfert galaxies.  The evolution of
low-luminosity active nuclei might be very strong at low redshifts,
meaning from roughly z = 0.2 to z = 0.4, as indicated in the
luminosity function at M$_B$ $\sim$ -23.0 summarized by Hartwick and
Schade (1990).  At face value, the density of such objects increases
from $z = 0.2$ to 0.4 by more than an order of magnitude. However,
the sample defining the z $<$ 0.2 luminosity function and the sample
defining the z $>$ 0.4 luminosity function were constructed in
different ways, and a wide variety of observational biases can be
imagined.  A single survey that spanned this range of redshifts would
be a valuable check.

For example, since active nuclei can often be detected in the 
X-ray band, one can investigate the evolution of an optically 
identified sample defined by X-ray flux.  It happens that the 
Einstein Observatory Extended Medium Sensitivity Survey 
(EMSS) sample of active nuclei (Della Ceca \etal 1994) spans 
the same redshift range and optical luminosity range 
addressed here, namely 0.1 $<$ z $<$ 1.0 and -21 $<$ M$_B$ $<$ -25, 
approximately.  For this sample, there is little if any 
evolution in the X-ray luminosity function, comparing a 
subsample with z $<$ 0.3 with a subsample with 0.3 $<$ z $<$ 0.8 
(Della Ceca \etal 1994). 

In order to be sure that the result for active nuclei detected in
X-rays can be generalized to optically selected active nuclei, we need
to know what fraction of the optically selected active nuclei are
X-ray sources.  Shanks \etal  (1991) studied a field imaged by ROSAT
to a limit in X-ray flux ten times fainter than that of the EMSS,
which yielded quasars with a median redshift of 1.5, essentially
identical to that of the quasars in T89/T94; even the shape of the
distribution at low redshifts is similar. They found that of the known
quasars in the field (selected to have z $<$ 2.2), 75\% of them
brighter than $B = 21.2$ were detected by ROSAT.  Because of this high
fraction, we would not expect that there will be a large number of
undiscovered Seyfert 1 galaxies in our field, since otherwise Shanks
\etal (1991) would have detected an excess of low-redshift sources in
their field.

The ability to detect a variable nucleus in a host galaxy depends on a
variety of factors, some of which relate to intrinsic properties of
the nucleus and host, and some of which relate to observational
aspects such the effects of image blur by seeing and the influence of
redshift on surface brightness detection thresholds.  At a fixed
redshift, if the nucleus is too luminous, the source will not qualify
as a Seyfert 1 galaxy, and if the nucleus is too faint, it will not be
detected to be variable.  Even for a fixed ratio of nuclear to
host-galaxy light, the detectability of variability will depend on the
profile type of the host galaxy: if the light from the host galaxy is
relatively extended, it will be easier to isolate the light from the
nucleus.  Yet another complication is color effects, since the
non-thermal nucleus will be bluer than the surrounding star light, and
the contrast in a fixed observer's band will depend on
redshift. (Clearly, those nuclei discovered in an optical survey are
those that do not suffer large extinction due to dust internal to the
host galaxy.)

In this study we devise an operational approach for detecting
variability in a sample of extended objects, meaning that there is no
special attempt to counteract the selection effects mentioned
above.  Rather, we develop an empirical procedure that yields light
curves for extended objects based essentially on considerations of the
signal-to-noise ratio in a single band (the photographic $J$, or$B_J$
band, where $B \sim J + 0.1$). In so doing, we can at least put lower
limits on the space density of low- luminosity active nuclei at
intermediate redshifts in a flux regime not previously explored.

The procedure may find other things, such as galaxies that appear to
be variable because of a projected quasar, variable star, or
supernova. Projected quasars would tend to have redshifts much higher
than plausible for normal host galaxies (and in any case we are
interested in finding them). According to T94, variable stars
selected in this field at these magnitudes are frequently red; hence a
projected star could be recognized by a measured color
uncharacteristic of a Seyfert galaxy. The variable stars in $SA$ 57
also tend to have amplitudes generally lower than the quasars, and
when diluted with light from the galaxy image onto which a star was
projected, the amplitude would be lower still. In our survey,
candidate supernovae would appear only at one epoch, but, as we
discuss below, our variability measurement and selection algorithms
are biased against finding such events.

\section{Procedures}

The data and basic procedures are described in detail in T89 and T94.
In T89 we considered a catalogue of sources based on plate MPF 1053,
which serves as a fiducial epoch against which other plates are
referenced.  Measurements were obtained on 9 plates over a baseline of
11 years.  The plates are homogeneous: all are IIIa-J plates obtained
at the prime focus of the Mayall 4-m telescope at Kitt Peak National
Observatory.  In T89, we established a sample of sources that were 1)
nominally stellar and 2) matched with the photometric catalogue of Koo
(1986).  The latter condition ensures that each source has multi-band
($UJFN$) photometry.  After applying a further condition that the
sources should have a core magnitude (called m$_1$) brighter than 24.0
(which yields total magnitudes for stellar sources brighter than about
22.5), the sample analyzed there amounted to 694 sources.  In this
study, we consider instead a sample that is complementary: it is also
matched with the Koo (1986) catalogue, but includes none of the 694
sources studied in T89 and T94.

The photometry consists of aperture magnitudes called m$_1$, m$_2$,
m$_3$, and m$_4$, according to the size of the aperture.  m$_1$ is the
core aperture (central 9 pixels, 0.279 arcsec per pixel), and m$_4$
corresponds to an aperture with a radius of 1.1 arcsec.  The index
m$_1$ - m$_3$ measures the image concentration.

Since the sources have various degrees of angular extent, the
magnitude cut is a sliding function of apparent size in order that the
limiting signal-to-noise ratio be approximately constant.
Specifically, for m$_1$ - m$_3$ $<$ 0.95, the magnitude cut is m$_2$
$<$ 23.5, and for m$_1$ - m$_3$ $>$ 0.95, the magnitude cut is m$_2$
$<$ 23.5 + 2.5 [(m$_1$ - m$_3$) - 0.95].  For a variety of reasons,
our techniques measure bright objects poorly, and we have accordingly
further eliminated all objects with m$_4$ $<$ 20.  This resulting
sample consists of 1636 objects in 1022 square arc minutes.
\footnote{In practice, our analysis is conducted on a sample of 2540
sources which satisfy our magnitude cuts. This includes the 1636
sources considered here, the 694 sources studied in T89 and T94, and
210 sources that are not matched with the catalogue of Koo (1986).
Source detection for Koo's catalogue used multiple plates but did not
include MPF 1053. Of the unmatched sources, roughly half (107) were
found to be on the border of regions excluded from the catalogues of
Koo (1986); the remaining unmatched sources were found to have
uncharacteristically small m$_1$ - m$_3$ or large variability.  Visual
inspection of the brightest of these sources on plate prints revealed
they were plate flaws. 'Unmatched' sources were excluded from the
final list of variable candidates.}

The aperture that gives the most robust measure of the magnitude is a
function of both image shape and magnitude, for example m$_1$ - m$_3$
and m$_4$. We estimated the values of these so-called optimum
apertures empirically by examining the scatter in independent measures
of magnitude as a function of aperture size, for samples defined in
different bins of magnitude and image concentration.  One would expect
different plates to have different optimum apertures for the same
object, but we found that this extra complication was not warranted:
each object is measured with the same aperture on all plates.  One
difference with respect to the analysis of T89/T94 is that the
normalization of the plate-to-plate differences of optimum magnitudes
($\Delta$m) is done by fitting a plane to the differences in the space
of m$_4$ and m$_1$ - m$_3$, instead of fitting line segments to the
differences versus m$_4$.  Table 1 gives specific values for the
adopted apertures.  After averaging measures at a common epoch and
after weighting the plates by their quality, we derive a measure of
variability called $\sigma$* for each source.  This is essentially the
r.m.s. value, after a variety of scale corrections have been applied
to ensure that the magnitude system on average matches that of plate
MPF 1053 (see T89 and T94 for details).

To provide a statistically consistent variability threshold, we
constructed a normalized quantity $$\sigma^*_N = (\sigma^* -
<\!\sigma^*\!>) \ / \ \sigma(\sigma^*),$$ where the mean and standard
deviation of $\sigma$* [$<\!\sigma^*\!>$ and $\sigma(\sigma$*),
respectively] are measured as a function of m$_4$ and m$_1$ - m$_3$.
The bivariate distribution of $\sigma$* in m$_4$ and m$_1$ - m$_3$
varies little as a function of m$_1$ - m$_3$ at a given m$_4$ except
at the threshold between stellar and non-stellar objects (m$_1$ -
m$_3$ = 1.05).  Hence, we simply divided the sample into two bins of
m$_1$ - m$_3$ ($<$ 1.05 and $\geq$1.05), and found $<\!\sigma^*\!>$
and $\sigma(\sigma$*) independently for these two samples.  To find
the mean $\sigma$*, $<\!\sigma^*\!>$, as a function of m$_4$, we fit a
low-order spline function and rejected individual points from the fit
if they were more than 3 standard deviations above or below the best
fit. This rejection process was iterated three times. The standard
deviation about this fit, $\sigma(\sigma$*), was then redetermined
more precisely using a sliding window in m$_4$ about each object,
adjusted to be wide enough to include a minimum of 100 objects, up to
a maximum width of 0.5 mag. Again, objects above 3 standard
deviations from the original fitting process were excluded from the
calculation.  To ensure that the standard deviation changed smoothly
with m$_4$, we fit a low-order spline to the individual standard
deviation measurements as a function of m$_4$.  Table 2 summarizes the
values of the mean and standard deviation of $\sigma$* as a function
of m$_4$ and m$_1$ - m$_3$ that we used in this analysis to construct
$\sigma$*$_N$.

We note that while the present sample of sources is supposed to have
extended image structure, it most likely contains some bona fide
stars, especially at the faint end.  For example, an object with 
m$_1$ - m$_3$ $<$ 1.05 but with m$_1$ $>$ 24.0 would not be included in the
source list of T89, but it will appear in the complementary list
considered here. A few such objects will have measured m$_4$ $<$ 22.5,
bright enough to be reliably measured for variability.

The criterion for variability that we adopt is simple: for m$_4$ $<$
22.5, the normalized $\sigma$*, $\sigma$*$_N$, must be larger than
2.5.  For comparison, we consider also a secondary sample consisting
of 35 sources with m$_4$ $>$ 22.5 and $\sigma$*$_N$ larger than 3.5.
The 14 sources in the first list can be considered to be prime
candidates, and are listed in Table 3. Two sources in the second list
are also included in Table 3 in order to define a sample complete to
$J < 22$, as defined by the total magnitudes from Koo (1986). These
are discussed in Section 4. All of the sources in Table 3 have
extended light profiles as defined both by the criterion
m$_1-$m$_3>1.05$ and by the r$_{-2}$ parameter of Kron (1980). We
inspected the images of each of these visually to check for any
peculiarities (such as close neighbors) that potentially might
influence the photometry.  No clear problems were found, but some
close neighbors are indicated in the notes to Table 3.

Some cases of candidate variables were matched with more than one
object in the Koo (1986) list.  Upon inspection, in all cases the
present catalogue misses a close object, and it is impossible to
recover which particular object was actually measured. It seems very
likely that the proximity of two or more objects has resulted in
photometric errors that appear as large $\sigma$*. We have deleted
from further consideration all cases where there were multiple
matches.  All of the close neighbors indicated in Table 3 were
substantially fainter than the ones resulting in multiple matches.

\section{Findings}

Figure 1 shows a plot of $\sigma$* versus m$_4$ for the sample of 1636
objects.  The format of this plot is directly analogous to that of
Figure 1 of T89.  Note that the noise level defined by the majority of
non-variable objects is respectably low, and in fact compares well
with that for stellar images (i.e. Figure 1 of T89), despite the
expected greater difficulty of measuring images with a variety of
profile types.  The upper envelope of $\sigma$* is quite well- defined
for the non-variable objects down to about m$_4$ = 22.5.

The candidate variable objects selected according to the criteria
mentioned in the previous section are marked. (Note that at faint
magnitudes there is an overlap between the marked and unmarked
objects.  This occurs because the compact images are normalized
differently from the extended images.)  The distribution of $\sigma$*
for the candidate variables is similar to that of the quasars
discussed in T89. In the latter case we had substantial spectroscopic
confirmation that the variables really are quasars (and therefore are
likely to be truly variable!). In the present case, the spectroscopic
confirmation is not as complete.

\subsection{Spectroscopic confirmation}

Spectra for 12 of the 16 candidates in Table 3 have have been
collected over the past decade using KPNO multi-fiber spectrographs on
the Mayall 4m telescope (Figure 2a) and more recently on the WIYN 3.5m
telescope (Figure 2b).\footnote{The WIYN Observatory is a joint
facility of the University of Wisconsin-Madison, Indiana University,
Yale University, and the National Optical Astronomy Observatories.}
These spectra were obtained as part of larger redshift survey programs
(KK88, Kron \etal 1991, Munn \etal 1997, Bershady \etal 1997).

Three candidates in Figure 2a clearly show evidence for the presence
of AGN: MgII is detected in Nser 100681 and 114264, and broad
H$\alpha$ emission is seen in Nser 110195. For Nser 100681, the MgII
line (as well as H$\beta$ and [O~III] $\lambda 4959,5007$) have been
independently confirmed from a spectrum taken with the KPNO 2.1m and
GoldCam spectrograph (Bershady, Trevese, and Kron 1991).  For Nser
114264, the MgII identification is less secure because it is so near
the blue limit of the spectral range, although it does avoid the
telluric absorption feature at 3700 \AA \ and it is at the correct
wavelength for the redshift determined from other lines. Notice also
the unusual combination of a strong break at 4000 \AA \ {\it and} the
presence of [O~III] $\lambda 4959,5007$ \ emission (H$\beta$ suffers
from being on a sky line). The last object in Figure 2a, Nser 110459,
has a secure redshift but no detected line emission; there is
insufficient evidence for the presence (or absence) of AGN.  In the
last two cases, increased signal-to-noise and extension of spectral
coverage to the blue or red (to cover either MgII or H$\alpha$) would
further secure identification of nuclear activity. The spectrum for
one additional object (NSER 116720) is not shown because it has
inadequate signal-to-noise to be of value.

Of the more recent spectroscopy shown in Figure 2b, only one
candidate's spectrum (Nser 105334) presents compelling evidence for
nuclear activity based on line ratios of [O~II] $\lambda 3727$,
H$\beta$ and [O~III] $\lambda 5007$.  We estimate log( [O~III]
$\lambda 5007$/H$\beta$ ) $\sim$ 0.76 $\pm$ 0.04 and log( [O~II]
$\lambda 3727$/H$\beta$ $\sim$ 0.41 $\pm$ 0.02, where the
uncertainties include random errors only. For example, we have made no
correction for H$\beta$ absorption. Also, the spectrum is
uncalibrated. To estimate the [O~II] $\lambda 3727$/H$\beta$ ratio, we
have used nominal throughput values as a function of wavelength,
calculated specifically for the adopted instrumental configuration.
The estimated values are just on the border of the Seyfert 2 region
defined by Tresse \etal (1996 and references therein), in a region
where some confirmed Seyfert 2's are found.

One other object in Figure 2b has a secure spectroscopic redshift:
Nser 108853.  While line emission is evident ([O~II] $\lambda 3727$
and possibly [O~III] $\lambda 5007$), the spectrum does not cover
H$\alpha$ or MgII. The other four objects in Figure 2b also show line
emission, however either there is only one line (identified as [O~II]
$\lambda 3727$ for Nser 113571, [O~III] $\lambda 5007$ for Nser 107772
and H$\alpha$ for Nser 118640), or the line is near a strong sky line
(Nser 106596). For the two cases where either MgII or H$\alpha$ is
supposed to be within the spectral range, the redshifts are uncertain.

Lastly, there is one object (Nser 104326) for which we have obtained
two spectra, $\sim$6 years apart (Figure 2d). Both spectra were
observed with identical apertures (3 arcsec diameter fibers). While
the first spectrum shows only one probable emission line ([O~II]
$\lambda 3727$) the second spectrum shows strong (and narrow) [O~II]
$\lambda 3727$, and [O~III] $\lambda 4959,5007$ at $z = 0.215$. The
first spectrum also has a very blue (rising) continuum below 5500 \AA.
While the second spectrum is not flux calibrated, it shows a strong
(Balmer) break not present in the first, and a redder continuum below
5500 \AA; these differences are not attributable to the lack of
calibration in the more recent spectrum. This object exhibits spectral
variability which we identify as Seyfert-like activity.

In summary, three objects are definitely confirmed Seyfert 1's (Nser
100681, 110195, 104326), one object is likely a Seyfert 1 (Nser
114264), and a fifth (Nser 105334) has line ratios marginally
consistent with a Seyfert 2. We identify these five as 'confirmed'
AGN. Of the remaining six spectroscopically observed candidates, none
can be definitively ruled out as AGN on the basis of the available
spectra.

%
%

\subsection{Candidates at m$_4$ $<$ 22.5}

There are many possible reasons why an extended image might
erroneously be selected as a variable, and without additional
arguments, one does not know how seriously to take the remaining
unconfirmed candidates. First, we can ask whether the number of
candidates in Table 3 is reasonable given expectations for the
number of Seyfert 1 galaxies to this depth. Such an estimate was made
by R. Burg (1987), who predicted the counts of Seyfert 1 galaxies
based on the statistics of active nuclei seen in the CfA survey.
According to this estimate, 24 Seyfert 1 galaxies are expected in this
area to this depth. But, our survey is sensitive only to a particular
subset as outlined in Section I, and on this basis the actual number
of candidates seems to be broadly reasonable.  Moreover, as mentioned
earlier, the results ROSAT identifications by Shanks \etal (1991)
constrain the population of active nuclei at B $\sim$ 21, with similar
conclusions.

One possible source of 'contamination' in our sample may be
supernovae. A supernova explosion within a galaxy could sufficiently
alter the galaxy's luminosity to give it a variability index that
would fall above our selection threshold. In this case, however, one
would expect the light curve to change (brighten) only at one epoch.
The light curves of the candidate variables, shown in Figure 3,
indicate that only two objects plausibly fit this pattern (Nser 104326
and 114264). Inspection of the images of these sources at the
appropriate epochs show no supernova events. Moreover, for both of
these objects there is spectroscopic evidence for Seyfert-like
activity. In short, there is no compelling evidence that supernovae
are significant contributors to the variability of our candidates.
Indeed, we would not expect to find supernovae in our sample of
variables because the anticipated total number of supernovae events is
small,\footnote{Scaling from the results of Pain \etal 1996, who find
34$^{+24}_{-16}$ supernovae yr$^{-1}$ deg$^{-2}$ for $21.3<R<22.3$, we
estimate there should be 0 to 3 supernovae in our entire series of
plate images. The scaling accounts for smaller area (0.284 deg$^2$),
shallower depth (translating to roughly a factor of two lower galaxy
surface density), and a sampling of the equivalent of 0.3 yr (7
epochs, each spaced by $>$ 1 year, and estimating supernovae will be
bright enough to raise our variability index above threshold for 15
days).} and our variability algorithm is not optimized for supernovae:
the apertures used to measure variability enclose a small fraction of
the total apparent galaxy area in a central region where the light
profile is brightest (worst contrast), and our glitch-suppression
algorithm (see T89) favors throwing out single epochs of large
variability.  The absence of supernovae events in our sample therefore
should {\it not} be used to constrain supernovae rates in faint galaxy
samples. Our variability algorithm is well suited for finding on-going
nuclear variability of low to moderate amplitudes.

If the list of candidate variables indeed consists of mainly Seyfert 1
galaxies, we would expect the distribution of m$_1$ - m$_3$ indices to
show a tendency to small (compact) values, since the nucleus should
affect the overall profile. Figure 4 plots m$_1$ - m$_3$ versus m$_4$
to explore this possibility. The five objects confirmed
spectroscopically to have active nuclei indeed have smaller m$_1$ -
m$_3$ than the parent sample, with the exception perhaps of Nser
104326. In general, the candidates at m$_4$ $<$ 21.7 appear to be
generally compact, whereas the fainter candidates (21.7 $<$ m$_4$ $<$
22.5) are not clearly distinct from the parent sample in this index.
This could suggest either that the lists of candidates becomes less
reliable with increasing faintness, or that the errors in m$_1$ -
m$_3$ grow with increasing m$_4$ (but the reliability remains high),
or that the fainter Seyferts are, in fact, more extended. Without
further information, it is difficult to distinguish between these
possibilities.

As previously mentioned, active nuclei are expected to be bluer than
the host stellar population, especially in a color index like $U-B$.
We therefore check the color distribution.  We have taken the $U-J$
color index from the Koo (1986) catalogue, and plot it in Figure 5
versus m$_4$.  The distribution of the candidate variables in this
plot with respect to the parent population follows very much the same
trends as noted for Figure 4, namely: The five objects confirmed
spectroscopically to have active nuclei (including Nser 104326)
clearly fulfill the expectation of bluer $U-J$ colors. The candidates
at m$_4$ $<$ 21.7 are somewhat bluer than the rest of the population,
and the fainter one (21.7 $<$ m$_4$ $<$ 22.5) are typical of the
population.

The bivariate distribution in $U-J$ color and m$_1$ - m$_3$ image
compactness indices, illustrated in Figure 6(a), also reveals the
m$_4$ $<$ 22.5 candidates are distinct from the rest of the
population. Here, however, even the 21.7 $<$ m$_4$ $<$ 22.5 candidates
are distinct. It appears that the color distribution for the candidate
variables is more clearly distinct from the parent population than the
size distribution. Note in particular the absence on any candidates in
the clump of points centered at roughly $U-J \sim 0.8$ and m$_1$ -
m$_3$ $\sim$ 1.15. This clump corresponds to normal galaxies with red
colors and large bulges such as types E and Sa.

A priori, it is not obvious how, or if the light curves of the
variable candidates will show correlations with other observables.
The light curves for the five reddest candidates ($U-J>0$) all show
large amplitude (0.5 mag) changes between epochs. Four of these are
the faintest at the same epoch. Inspection of the spatial locations of
these objects, however, shows they are randomly positioned in the
field. Therefore this coincidence is not likely the result of a plate
flaw. One of the bluest objects also is faintest at this epoch and has
a comparably large amplitude change. Hence there is no strong
indication that the light curve amplitude and shape correlate with
color. The shapes of the light curves also show no obvious trends with
either m$_4$ or m$_1$ - m$_3$.

Finally, we can use the spectroscopic redshift information to explore
whether the candidates are unusually distributed in luminosity or
redshift. The Hubble diagram in Figure 7 compares the 14 variables
candidates with spectroscopically confirmed galaxies (Munn \etal 1997)
and QSOs (Kron \etal 1991) in the same field. For candidates without
spectroscopic redshifts, photometric redshifts are calculated using a
method similar to that described by Connolly \etal (1995).

The five objects confirmed spectroscopically to have active nuclei
span a large range in luminosity. The apparently brightest candidates,
Nser 100681, is also the most luminous extended source in the field,
while Nser 105334 is roughly at 0.1 L$^*$.\footnote{We use L$^*$ to
refer to the normal galaxy luminosity function and corresponds to M$_B
\sim -21$ for H$_0=50$ km s$^{-1}$ Mpc$^{-1}$} One other source (Nser
106596) is exceptionally luminous, but its spectroscopic redshift is
uncertain. Were it placed at its photometric redshift, its luminosity
would be $\sim$L$^*$. Nonetheless, while the bulk of the candidates
appear typical in luminosity of the rest of the field galaxy
population, {\it the redshift distribution is skewed to higher values.}
With the exception of Nser 118640, the variable candidates are all at
or {\it above} the median redshift of the field galaxy sample ($z \sim
0.22$ for $B_J = 22.5$). Nser 118640 is itself in an unusual region of
the Hubble diagram.

The photometric redshifts in the above analysis are in principle prone
to systematics if the sources are dominated by AGN light.  Systematics
might arise because the ``training set'' used to calibrate the
photometric redshift relationship contains excludes AGN. In other
words, the mapping of redshift onto the space of $UJFN$ apparent
magnitudes is done for different types of spectral energy
distributions. Indeed, of the five confirmed AGN, the dispersion
between the spectroscopic and photometric redshift is 0.18 in
redshift, roughly 3.5 times larger than the average galaxy to $B_J<22$
using the same photometric data. On the other hand, the difference
between spectroscopic and photometric redshifts for the same five
objects is close to zero in the mean. At the very least, photometric
redshifts are useful here for identifying if the candidates have
unusual colors. It appears likely, however, that photometric redshifts
are reliable for these candidates on average, but unreliable for
individual targets.

Overall we conclude that Table 3 does contain many bona fide active
nuclei. This impression is based on the success of spectroscopically
identifying five objects; the bivariate distribution in color and
image structure; the distribution in redshift; and the sharp upper
envelope of points in Figure 1. At minimum, Table 3 can be considered
to be a good list for further spectroscopic investigation.

\subsection{Candidates at m$_4 > 22.5$}


The fainter candidates, m$_4$ $>$ 22.5, likely comprise a number of
truly stellar objects that were missed in our previous survey because
of the way we applied selection cuts to the parent sample. Indeed,
over half (19 out of 35) of these variable candidates have
m$_1-$m$_3<1.05$, and hence are consistent with stellar image
profiles. Of the 19 m$_4>22.5$ candidates with m$_1-$m$_3 < 1.05$,
however, about half (nine) are classified by Kron (1980) as
extended.\footnote{Kron classified image profiles on the basis of
  their second inverse moment, r$_{-2}$. In general r$_{-2}$ and
  m$_1-$m$_3$ are highly correlated. Different classifications
  based on these two indices reflect different sensitivities to a
  variety of image profiles, as well as noise. A comparison of r$_{-2}$
  to m$_1-$m$_3$ for comparably faint objects in this field can be
  found in Koo \etal (1986).} Yet of the 16 m$_4> 22.5$ candidates
with m$_1-$m$_3 > 1.05$, only two of these are classified by Kron
(1980) as stellar. Hence there is some indication of extended image
structure for over 90\% of this faint sample.

At the bright end of the m$_4 > 22.5$ candidate list, one object (Nser
17671) is independently selected as a QSO candidate from KKC (\#6,
m$_4 = 22.54$). This is an object classified as stellar by Kron
(1980), but extended based on a value of m$_1-$m$_3 = 1.134$.  Nser
17671 has a tentative spectroscopic identification as a narrow
emission-line galaxy at $z = 0.518$ based on a single line identified
as [O~II] $\lambda 3727$ (KK88, Table 1, \#73). The spectrum, however,
does not go blue enough to detect Mg II, or red enough to detect
H$\alpha$. Majewski \etal (1992) also find this object to be variable
and with no detectable proper motion.

At the faint limit of the m$_4 > 22.5$ candidate list, three other
objects are found in Majewski \etal's sample. Two have $B_J>23$.  All
three are stellar according to the m$_1$-m$_3$ criterion and Kron
(1980). None pass Majewski \etal's variability threshold, while two
have detectable proper motions (one of which is very large). However,
all three objects have $U-J > 1.75$ (one is not shown in Figure 6b
because it is a $U$ band dropout). Indeed, these are the only three
objects in the faint sample with such red colors, and they are likely
to be stars ({\it cf} Figure 3 of T94).

Figure 6(b) shows the m$_4$ $>$ 22.5 candidates in a plot similar to
Figure 6(a), which reveals a disproportionate number of candidates
with blue $U-J$ colors and compact image structure. This is expected
if the fainter sample has a higher fraction of blue stellar sources
which are bona fide quasars. The remaining variable candidates, with
m$_1-$m$_3 > 1.05$, are typically equally as blue as the stellar
candidates.  Without spectroscopic confirmation the reliability of
this fainter sample is unknown. In general, however, the variability
amplitudes of the fainter candidates (Figure 1) are consistent with
those of known quasars.

\section{The surface density of AGN to $B_J=22$}

We define $B_J$ operationally here as $J_K$, listed in Table 3,
because this magnitude is a good approximation to a total magnitude,
independent of image size and shape. A sample of extended variable
objects, complete to $B_J=22$, can be constructed by (i) adding two
fainter candidates (Nser 101951 and 107772) to our primary sampled
defined by m$_4 < 22.5$; but (ii) removing the faintest m$_4 < 22.5$
candidate with $B_J > 22$ (Nser 105334, a confirmed AGN). This
'complete' sample is useful for assessing the surface density of AGN:
at these depths the spectroscopic completeness of our QSO survey in
$SA$ 57 is relatively high (Kron \etal 1991), and other surveys
exist for comparison (e.g. Zitelli \etal 1992).

The two additional extended variable candidates are only $\sim 0.25$ mag
fainter than our primary sample limit in m$_4$, and all indications are that
these two objects are good candidates. For example, both are blue
($U-J<0$).  Their light curves (Figure 3) are also comparable to the
brighter candidates with the one exception that the 7th epoch is
excluded from the estimate of $\sigma$* for Nser 107772. (This
exclusion is mandated by a glitch-suppression algorithm, described
in T89.) Based on m$_1$-m$_3$, Nser 107772 is quite compact, with
$U-J$ comparable to the confirmed AGN. Yet while it would be
considered 'stellar' on the basis of m$_1$-m$_3$, it has a large r$_1$
and is classified as a galaxy in the catalogue of Koo (1986).  In
short, Nser 107772 appears to be large, blue galaxy with a compact core.
Both objects are among the apparently largest galaxies for their
apparent magnitude.

In total, there are 15 extended variable candidates to $B_J=22$, four
of which are spectroscopically confirmed AGN. In comparison, there are
26 confirmed QSOs (i.e. AGN with stellar image structure) in $SA$ 57
to the same limit. In addition, there are five {\it stellar} objects
satisfying high variability and lack of proper motion but without
spectroscopic confirmation (T94). Because all other {\it stellar}
objects meeting these variability and proper motion criteria are bona
fide QSOs, we assume here that these 5 objects are also QSOs.  Adding
our four confirmed extended variables to the 31 QSOs yields a total
surface density of 123 $\pm$ 21 deg$^{-2}$ in $SA$ 57.

Zitelli \etal (1992) find a similar surface density, 115 $\pm$ 16
deg$^{-2}$, to the same magnitude limit, and in a separate
field. Their sample is 100\% spectroscopically confirmed. Yet in the
limiting case where all of our extended variables are bona fide AGN,
the surface density in $SA$ 57 would reach 162 $\pm$ 24 deg$^{-2}$,
with extended sources contributing 33\%. Zitelli \etal use color and
variability selection strategies for stellar sources similar to our
own. There are, however, a few differences between the surveys. For
example, we have proper motion information, whereas their sample
includes selection by spectroscopic features seen on a grism plate.
One other difference is that they do not specifically target sources
with extended image structure.

Based on spectroscopic and photometric redshift information, the
majority of our extended variable candidates have luminosities less
than 10\% of the nominal demarkation between QSOs and AGN (M$_B = -23$,
H$_0=50$ km s$^{-1}$ Mpc$^{-1}$, q$_0=0.5$). This luminosity estimate
is an upper limit since the measured apparent magnitude is within a
large aperture that includes a non-negligible contribution of light
from the host galaxy. The surface density of {\it confirmed} M$_B >
-23$ AGN to $B_J = 22$, including 7 stellar sources and four extended
sources, is 39 $\pm$ 12 deg$^{-2}$, again in good agreement with
Zitelli \etal's value of 40 $\pm$ 9. In the limiting case where
all of our extended variable candidates are bona fide AGN, and we
include $5 \times \frac{7}{26}$ of the unconfirmed, non-moving stellar
variables (5 is the total number of unconfirmed, non-moving stellar
variables; $\frac{7}{26}$ is the fraction of confirmed QSOs with
M$_B>-23$), the surface density would reach 82 $\pm$ 17 deg$^{-2}$
for M$_B > -23$ with extended sources contributing 64\%. The majority
of the unconfirmed AGN are between $0.2<z<0.4$, whereas Zitelli \etal
find no AGN at $z<0.4$.

The total number of extended variable candidates corresponds to
$\sim$2\% of the surface-density of all extended sources to
$B_J<22$. These candidates vary at roughly $> 0.1$ mag.  In contrast,
Kochanski \etal (1996) find only 0.74\% of all sources vary by 0.026
mag or more to the same magnitude limit. However, T89 found $\sim$9\%
of the stellar sources vary at roughly $>$ 0.1 mag to $B_J<22.5$ (or
$\sim 5$\% to $B_J<22$). T89's result is roughly two orders of
magnitude higher than Kochanski \etal (by Kochanski \etal's estimate)
for stellar sources. Yet T89's sample of candidate stellar variables
independently identified spectroscopically confirmed QSOs with high
completeness and reliability. Similarly, we have spectroscopically
confirmed that at least one third of our extended variable candidates
are bona fide AGN. Further, as we have discussed in Section 3.2, the
number of extended variable candidates is comparable to the total
number of Seyfert 1's expected at this magnitude limit. We conclude,
therefore, that the surface density of variable sources we find is
likely reliable.

\section{Summary}

We have shown that it is possible to select AGN in sources with
extended light profiles on the basis of variability alone, in a manner
similar to that which we have successfully identified stellar AGN
(QSOs). A posteriori, several of our extended variable candidates have
been confirmed spectroscopically to be bona fide AGN. However, we
still lack adequate spectroscopy to confirm or rule our the existence
of nuclear activity for the majority of our sample. The color and size
distributions of the candidates shows that at brighter magnitudes the
sample is preferentially bluer and more compact than non-variable
galaxies. The confirmed AGN are in this magnitude range. At fainter
magnitudes, the sample is more representative of the entire population
of extended sources. The redshift distribution of the sample, however,
is skewed to values above the median for normal galaxies at the depth
of this survey. On these bases, and from our preliminary spectroscopic
success, we believe we have identified a sample worthy of further
spectroscopic investigation. If all of these candidates are bona fide
AGN, this represents a 30\% increase in the surface density of all AGN
(i.e. including QSOs) at $B_J = 22$. At the very least, AGN which
appear extended in ground-based images with $\sim$ 1-1.5 arcsec seeing
represent 10\% of the total AGN population.

\acknowledgements

We gratefully acknowledge Dave Andersen for assistance with reducing
WIYN/Hydra spectra and and estimating photometric redshifts;
Alessandro Bunone for photometric error estimates; Steve Majewski for
information from his proper-motion survey; and Jeff Munn and David Koo
for their efforts in producing the spectra from the Kitt Peak Galaxy
Redshift Survey. We thank Di Harmer and other members of the NOAO WIYN
Queue team for assistance in gathering spectroscopic data. MAB
acknowledges support from the Italian CNR; NASA through Graduate
Fellowship NGT-50677 and Hubble Fellowship HF-1028.01-92A from Space
Telescope Science Institute, which is operated by the Association of
universities for Research in Astronomy, Incorporated, under contract
NAS5-26555; Sigma Xi, Grants-in-Aid-of-Research; and research funds
from Penn State University. DT acknowledges support from MURST.

\clearpage

\begin{figure}
\plotfiddle{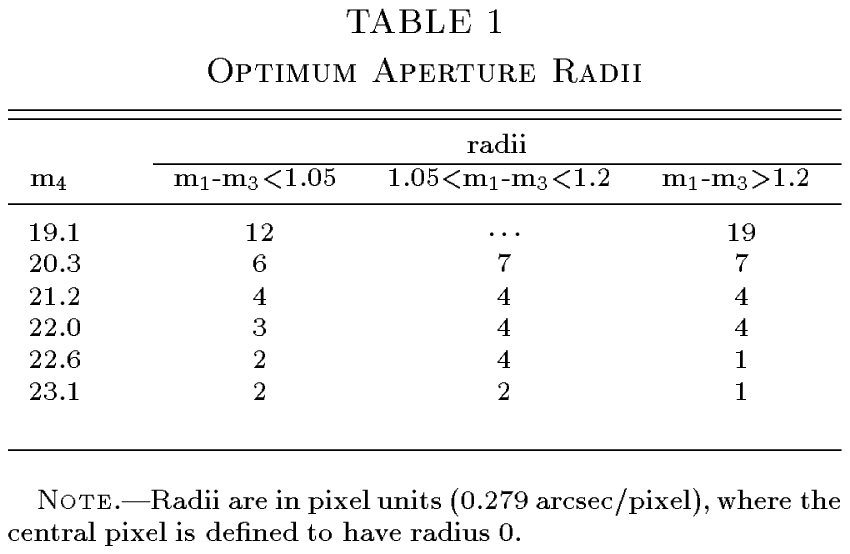}{8in}{0}{100}{100}{-325}{0}
\end{figure}
\begin{figure}
\plotfiddle{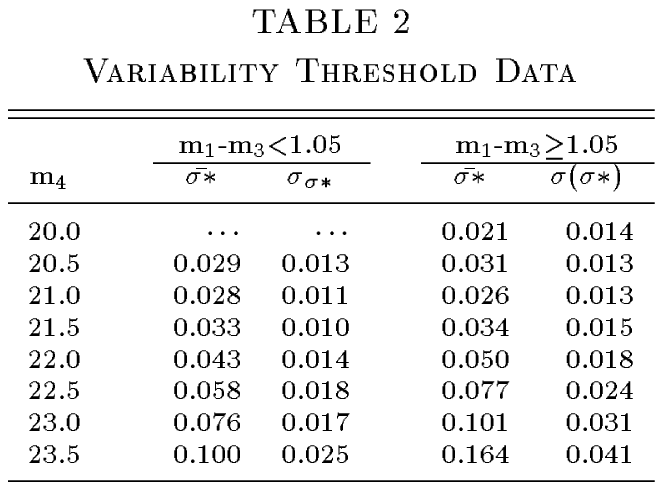}{8in}{0}{100}{100}{-325}{0}
\end{figure}
\begin{figure}
\plotfiddle{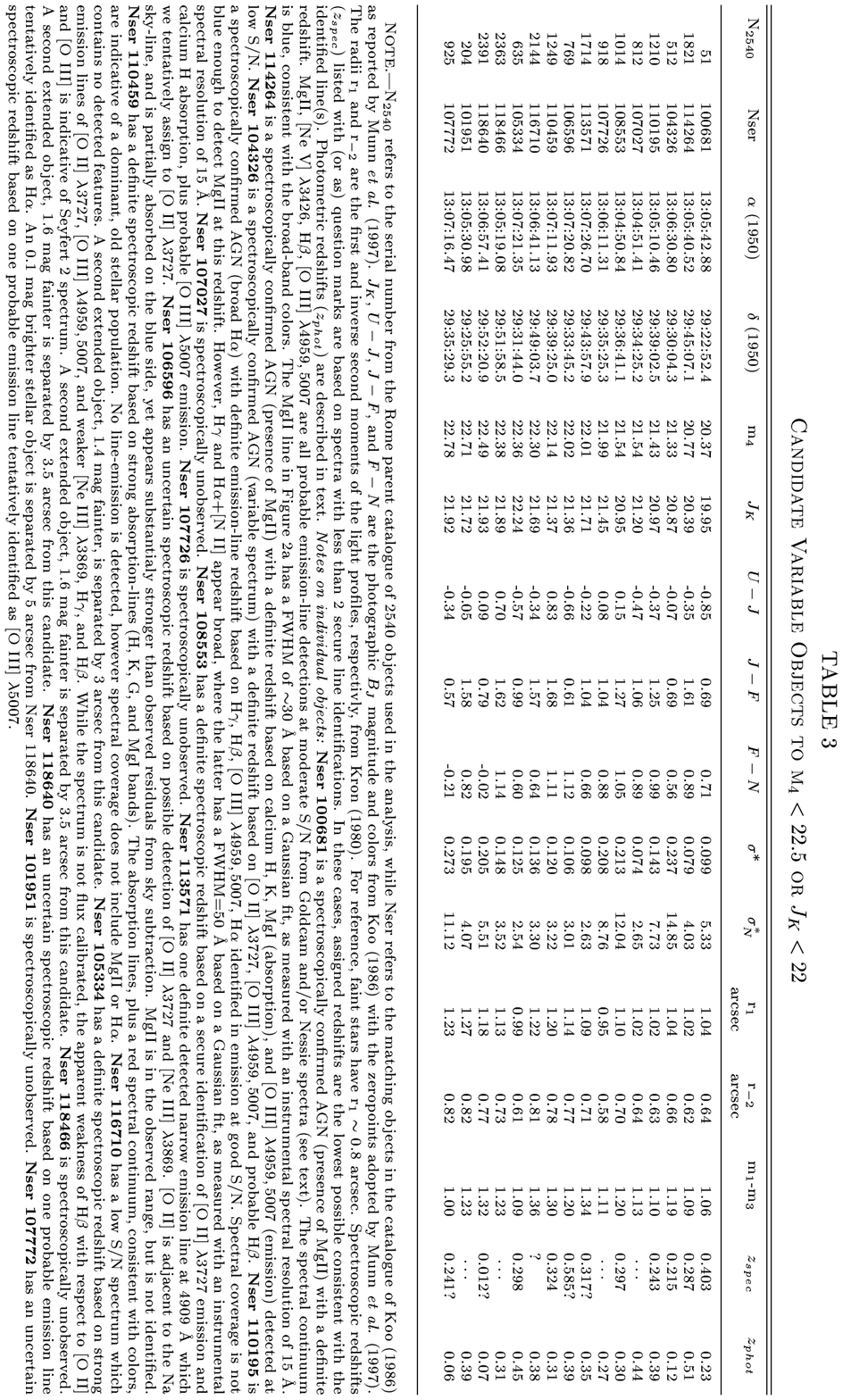}{8in}{180}{100}{100}{300}{675}
\end{figure}

\clearpage

\figcaption[]{Weighted root-mean-square deviation of the magnitude
  differences with respect to the mean for each object, over the 11
  year baseline, as a function of J magnitude on MPF 1053
  (m$_4$(1053)) for the 1636 objects in the survey. The variability
  criteria for compact (m$_1$-m$_3$$<$1.05) and diffuse sources
  (m$_1$-m$_3$$>$1.05) are indicated by the dashed and solid lines
  respectively. These are derived from the photometric errors (Table
  2). For m$_4$$<$22.4, a 2.5 $\sigma$ threshold is imposed (lower
  dashed and solid curves); for m$_4$$>$22.5, a 3.5 $\sigma$ threshold
  is imposed (upper dashed and solid curves). Variable candidates are
  specially marked in categories of: spectroscopically confirmed AGN,
  close to neighboring objects, or otherwise as a function of image
  compactness. Four unmarked objects above thresholds for m$_4<22.5$
  are multiply matched between two independent catalogues, as
  discussed in text. Arrows point to two candidates with m$_4>22.5$
  but $J_K<22$.
\label{fig1}}

\figcaption[]{(a) Spectra of four variable candidates from the KPNO 4m
telescope (instrumental configurations and exposures differ for each
object, as described in Munn \etal 1997). Three spectra confirm the
presence of AGN (Nser 100681, 114264, and 110195), as discussed in the
text. A fourth spectrum (Nser 110459) yields a reliable
redshift. Confirmed lines are detailed in the Notes to Table 3, while
positions of common emission/absorption lines are marked here at the
appropriate redshift for reference.  \label{fig2a}}

\setcounter{figure}{1}

\figcaption[]{(b) Spectra of six variable candidates from the WIYN
3.5m telescope and Hydra multi-fiber spectrograph (3 arcsec fibers,
400 lines/mm grating, spectral resolution [$\lambda$/$\Delta\lambda$]
of $\sim$780, three hours integration for Nser 113571, 106596, 105334,
six hours integration for Nser 107772, nine hours integration for Nser
108553, 118640). Two objects (Nser 108553 and 105334) have secure
redshifts based on strong multiple emission or absorption lines. For
Nser 105334, note the weakness of H$\beta$ relative to [O~III]
$\lambda 4959,5007$ and [O~II] $\lambda 3727$ \AA, which identifies
this source as a possible Seyfert 2. In all other cases there is
insufficient evidence for the presence of an AGN. Confirmed lines are
detailed in the Notes to Table 3, while positions of common
emission/absorption lines are marked here at the appropriate redshift
for reference. \label{fig2b}}

%

\setcounter{figure}{1}

\figcaption[]{(c) Spectra of Nser 104326 taken $\sim$6 years apart but
with the same aperture. (See caption to Figure 2a and 2b for
descriptions of observations.) The spectra show dramatically different
continuum shapes and emission-line equivalent widths.\label{fig2c}}

\figcaption[]{Light-curves for the 16 candidate variables in Table 3,
  ordered by m$_4$. Magnitudes (m), and their uncertainties (1 sigma)
  are defined within the optimum aperture (Table 1) and are plotted
  with respect to the average magnitude ($<\!m\!>$).\label{fig3}}

\figcaption[]{Image compactness (m$_1$-m$_3$) as a function of J
magnitude [m$_4$(1053)] for the same sample in Figure 1. Variable
candidates are specially marked in categories of: spectroscopically
confirmed AGN, close to neighboring objects, or otherwise as a
function of $U-J$ color. The dashed line at m$_1$-m$_3=1.05$ demarks
regions where objects are nominally stellar or extended. Arrows point
to two candidates with m$_4>22.5$ but $J_K<22$. \label{fig4}}

\figcaption[]{$U-J$ color as a function of $J$ magnitude [m$_4$(1053)]
for the same sample in Figure 1. Variable candidates are specially
marked in categories of: spectroscopically confirmed AGN, close to
neighboring objects, or otherwise as a function of image
compactness. Arrows point to two candidates with m$_4>22.5$ but
$J_K<22$.\label{fig5}}

\figcaption[]{(a) $U-J$ color as a function of image compactness
(m$_1$-m$_3$) for survey objects in the range
20$<$m$_4$$<$22.5. Variable candidates are specially marked in
categories of: spectroscopically confirmed AGN, close to neighboring
objects, or otherwise as a function of normalized $\sigma$*
($\sigma$*$_N$). The dashed line at m$_1$-m$_3=1.05$ demarks regions
where objects are nominally stellar or extended.\label{fig6a}}

\setcounter{figure}{5}

\figcaption[]{(b) $U-J$ color as a function of image compactness
(m$_1$-m$_3$) for survey objects in the range m$_4$$>$22.5. Variable
candidates are specially marked according to the normalized $\sigma$*
($\sigma$*$_N)$. The dashed line at m$_1$-m$_3=1.05$ demarks regions
where objects are nominally stellar or extended. Arrows point to two
candidates with m$_4>22.5$ but $J_K<22$.\label{fig6b}}

\figcaption[]{Apparent magnitude ($J_K$) vs. redshift for objects in
SA 57: spectroscopically confirmed QSOs (T94), extended galaxies with
secure spectroscopic redshifts (dots, Munn \etal 1997), and the 16
variable candidates from Table 3, here (see key). In cases where
spectroscopic redshifts ($z_{spec}$) are not available for the
variable candidates, photometric redshifts ($z_{phot}$) are used
instead. Lines of constant luminosity for M$_B=-23$ and -21 are drawn
assuming H$_0=50$ km s$^{-1}$ Mpc$^{-1}$, q$_0=0.5$ and
$\alpha=-1$. \label{fig7}}

\clearpage

\begin{figure}
\plotfiddle{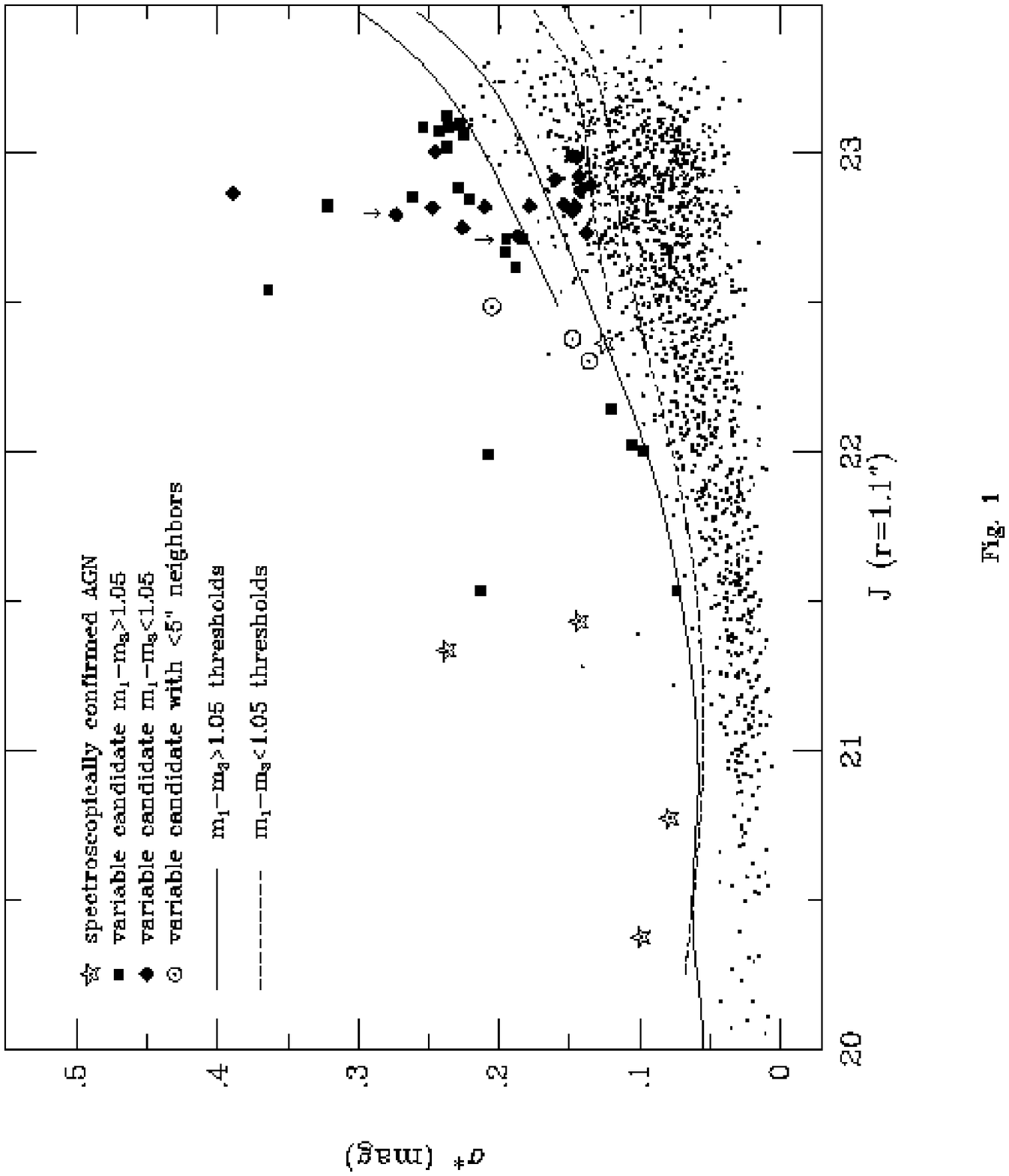}{8in}{0}{100}{100}{-315}{-75}
\end{figure}

\clearpage

\begin{figure}
\plotfiddle{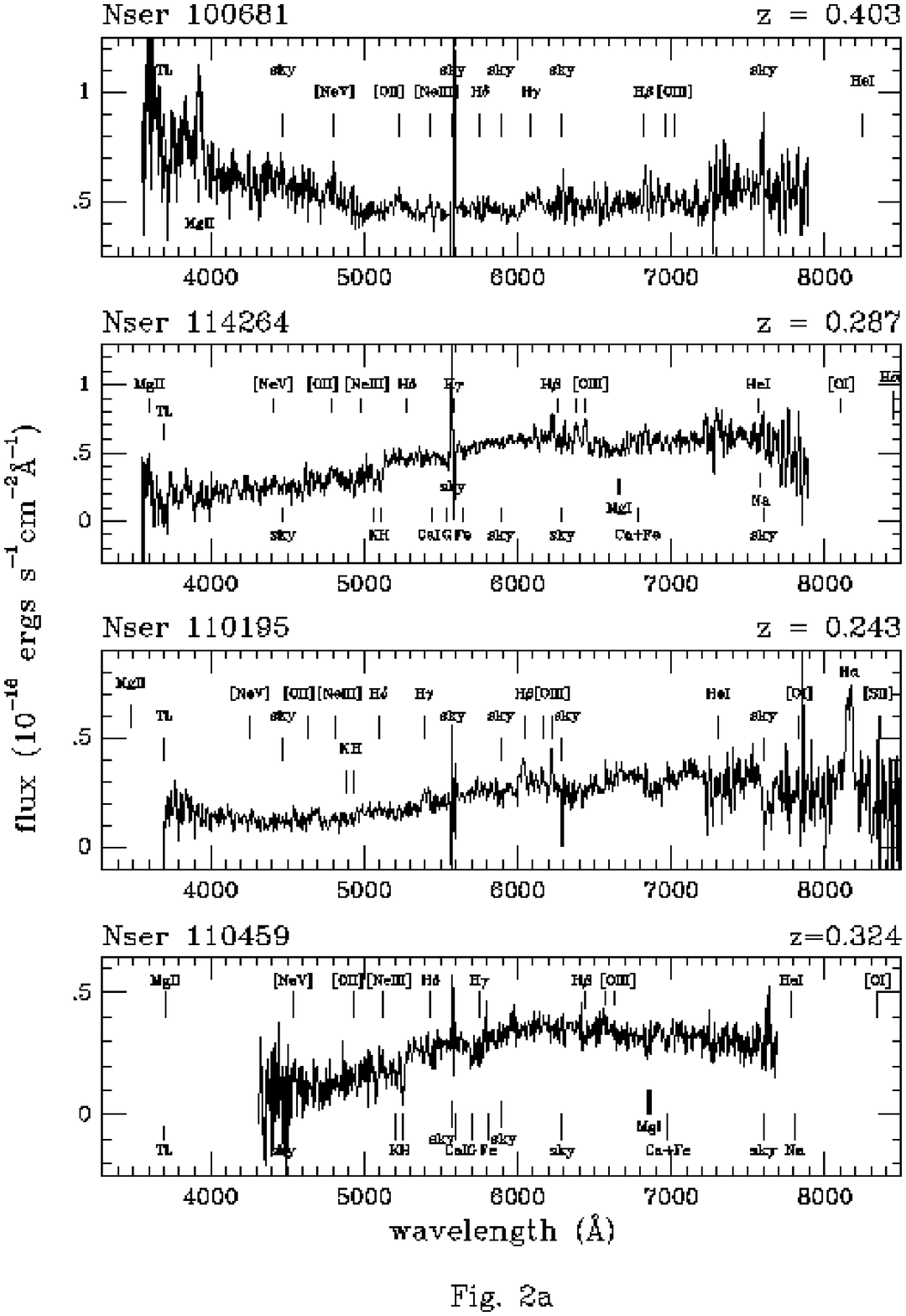}{8in}{0}{95}{95}{-280}{-100}
\end{figure}

\clearpage

\begin{figure}
\plotfiddle{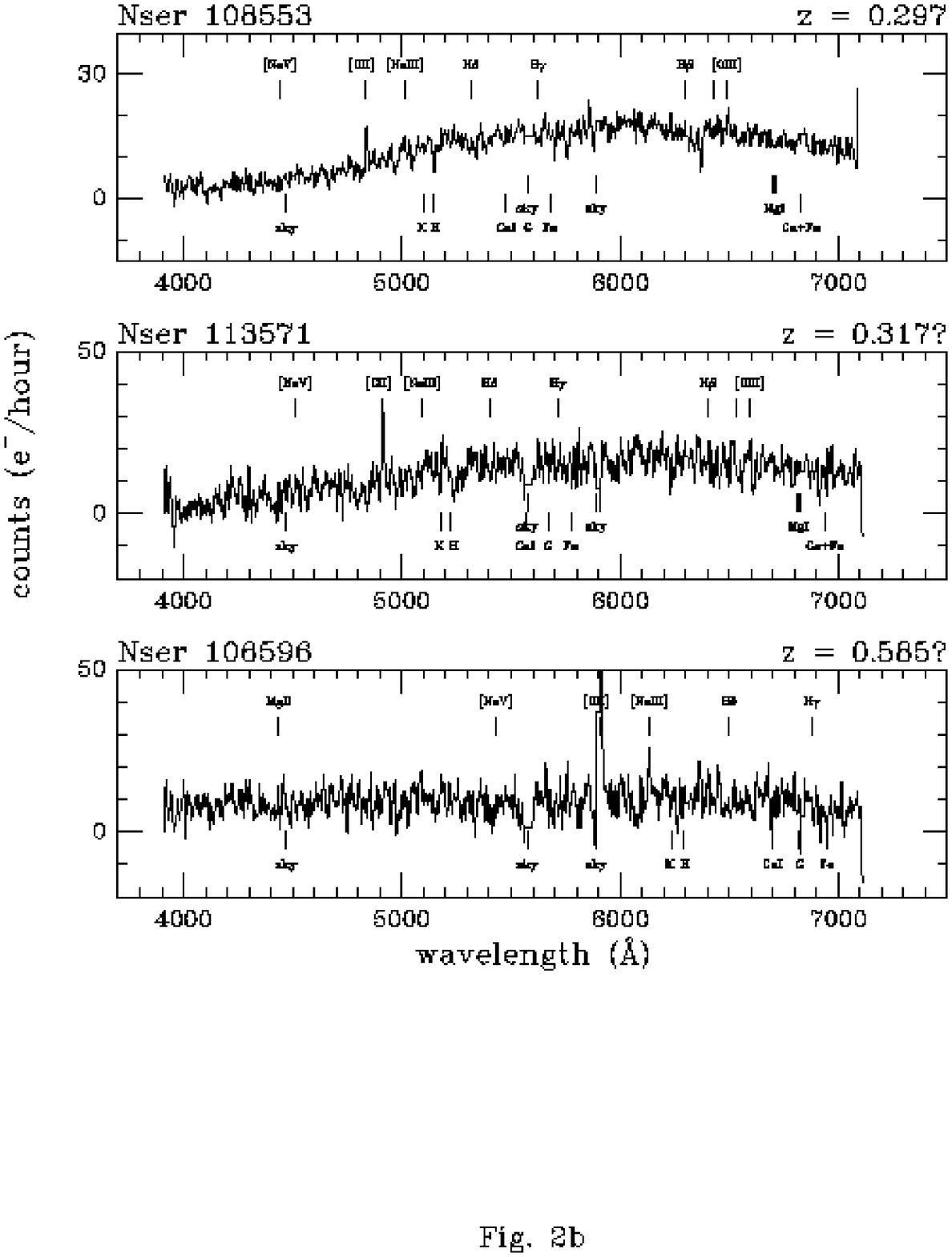}{8in}{0}{95}{95}{-280}{-120}
\end{figure}

\clearpage

\begin{figure}
\plotfiddle{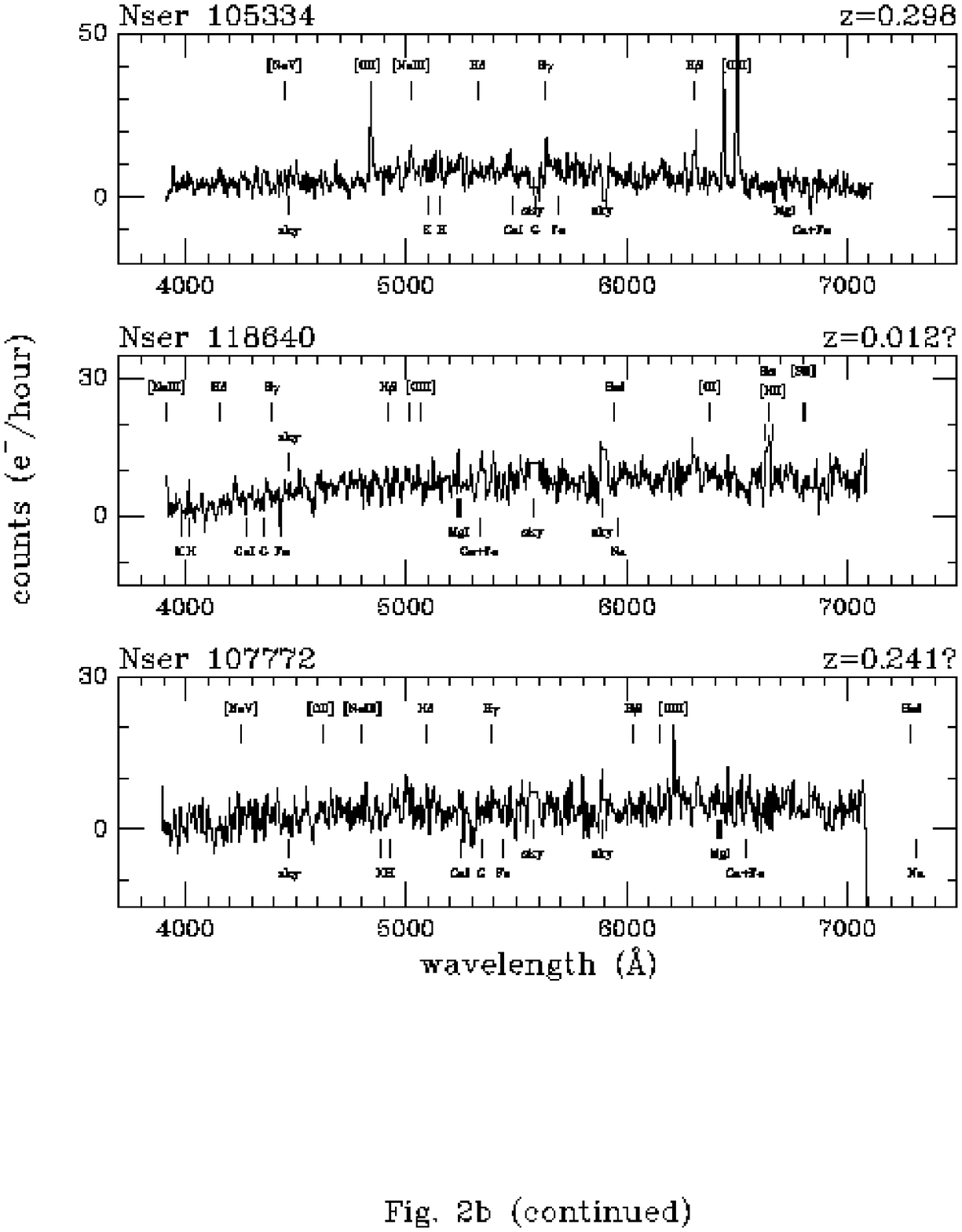}{8in}{0}{95}{95}{-280}{-120}
\end{figure}

\clearpage

\begin{figure}
\plotfiddle{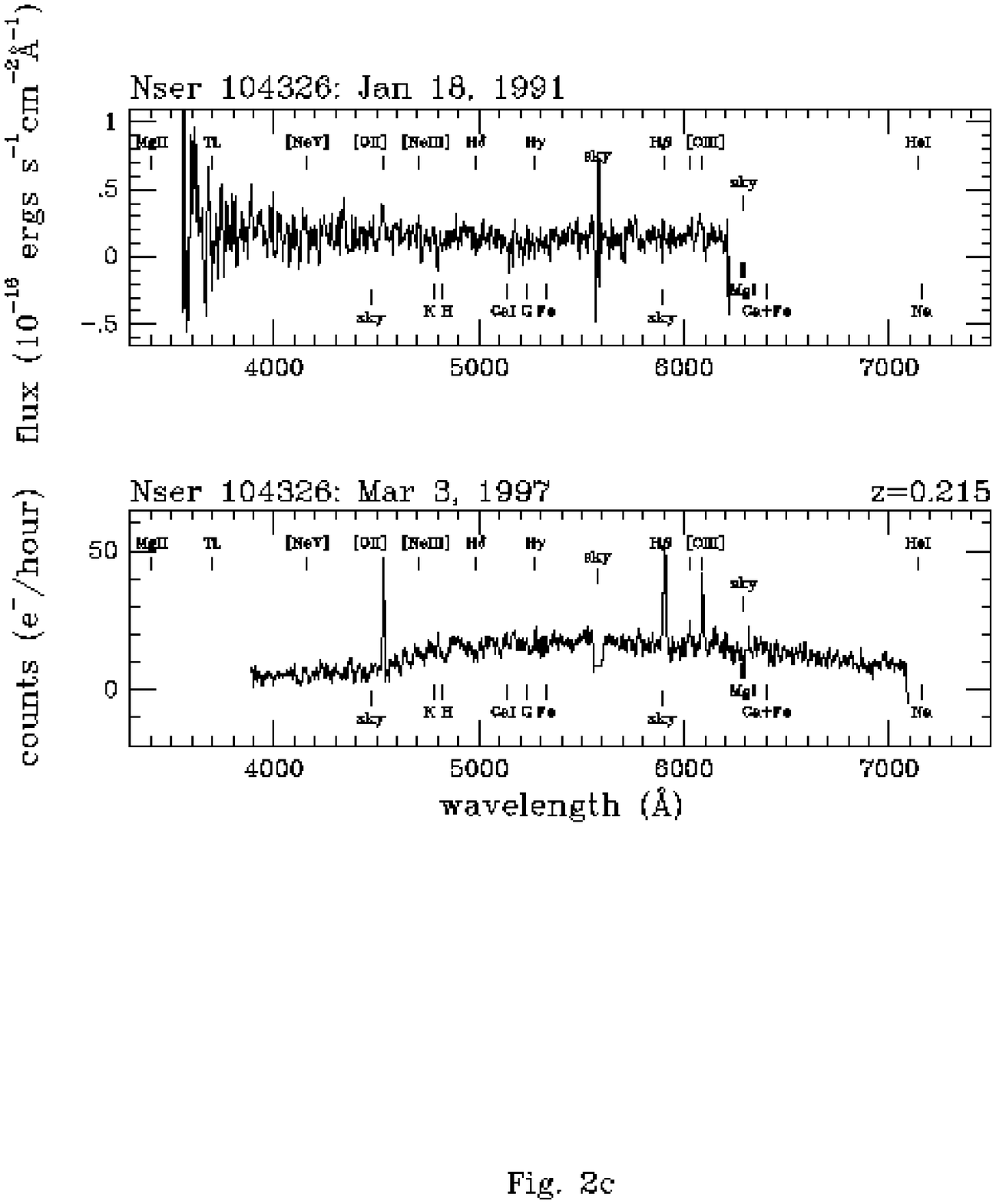}{8in}{0}{95}{95}{-280}{-140}
\end{figure}

%

\clearpage

\begin{figure}
\plotfiddle{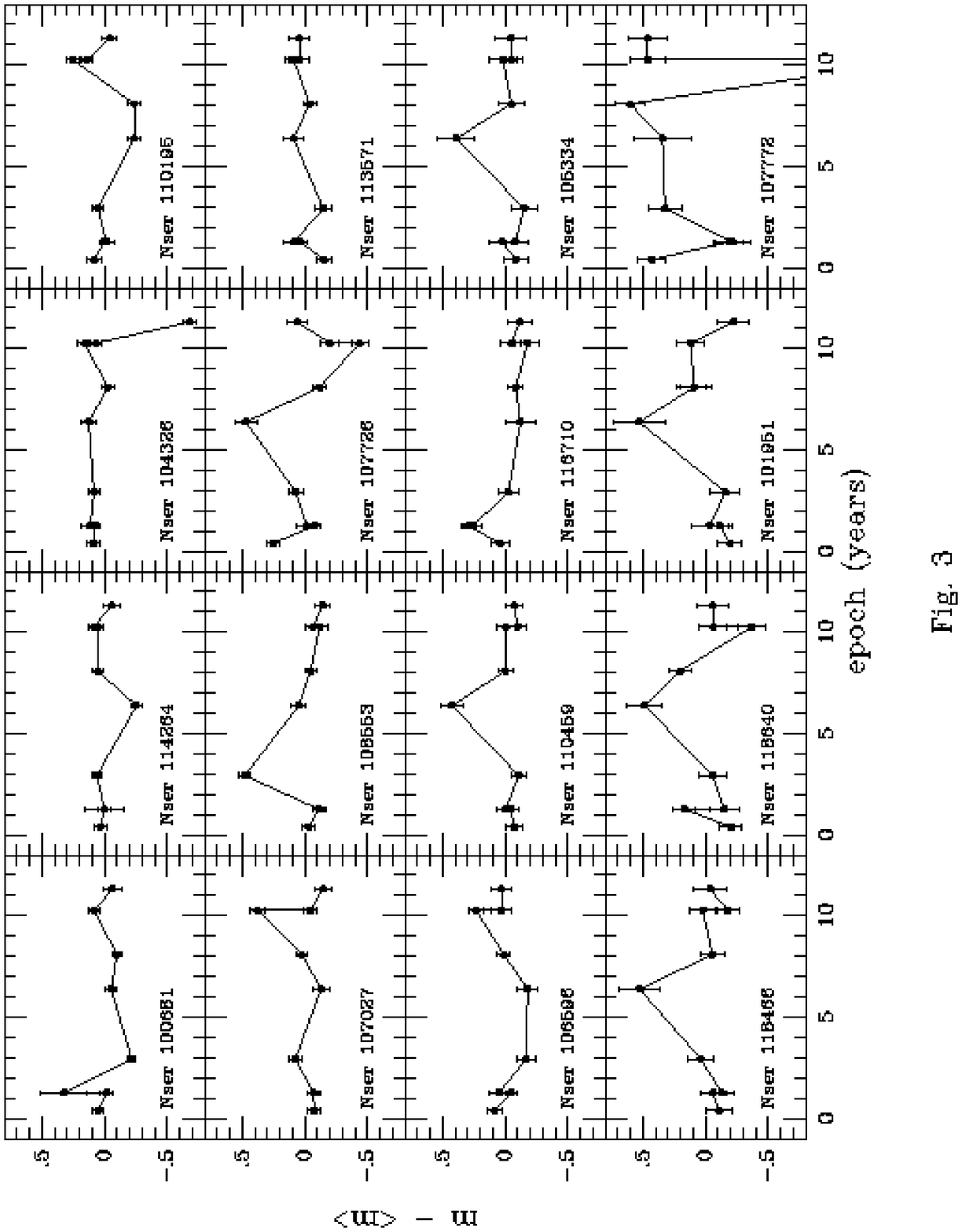}{8in}{0}{90}{90}{-285}{-65}
\end{figure}

\clearpage

\begin{figure}
\plotfiddle{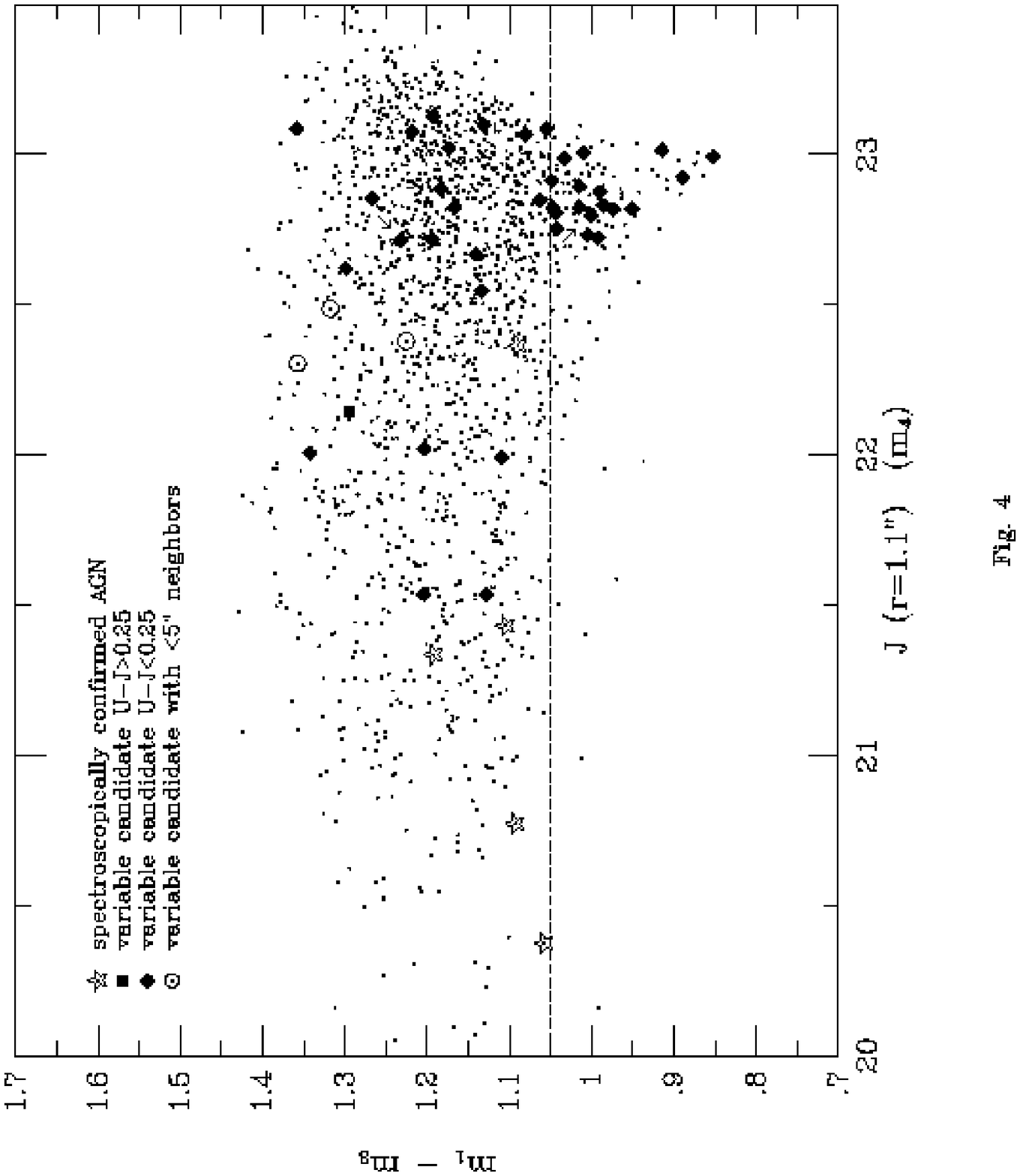}{8in}{0}{100}{100}{-315}{-75}
\end{figure}

\clearpage

\begin{figure}
\plotfiddle{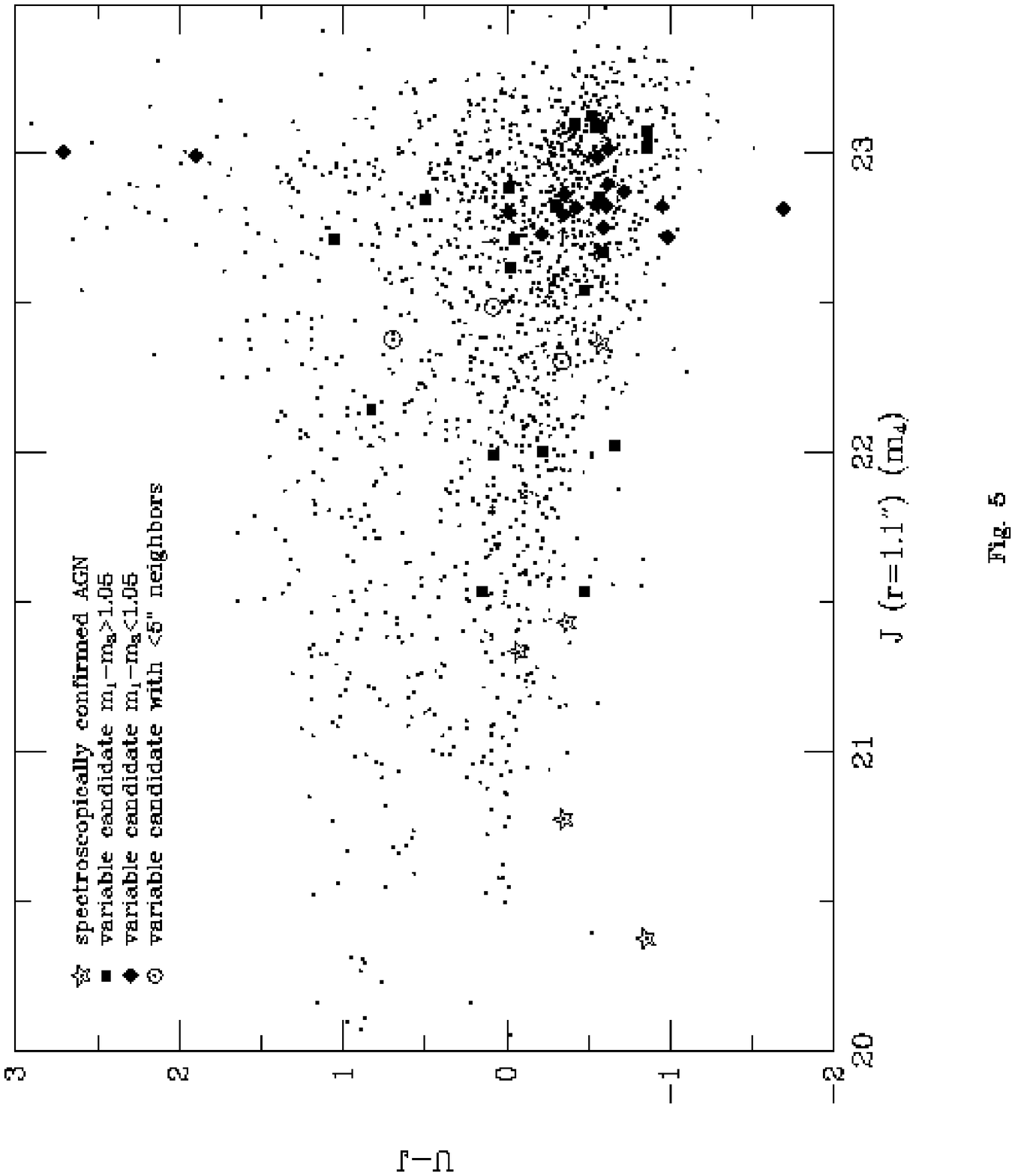}{8in}{0}{100}{100}{-315}{-75}
\end{figure}

\clearpage

\begin{figure}
\plotfiddle{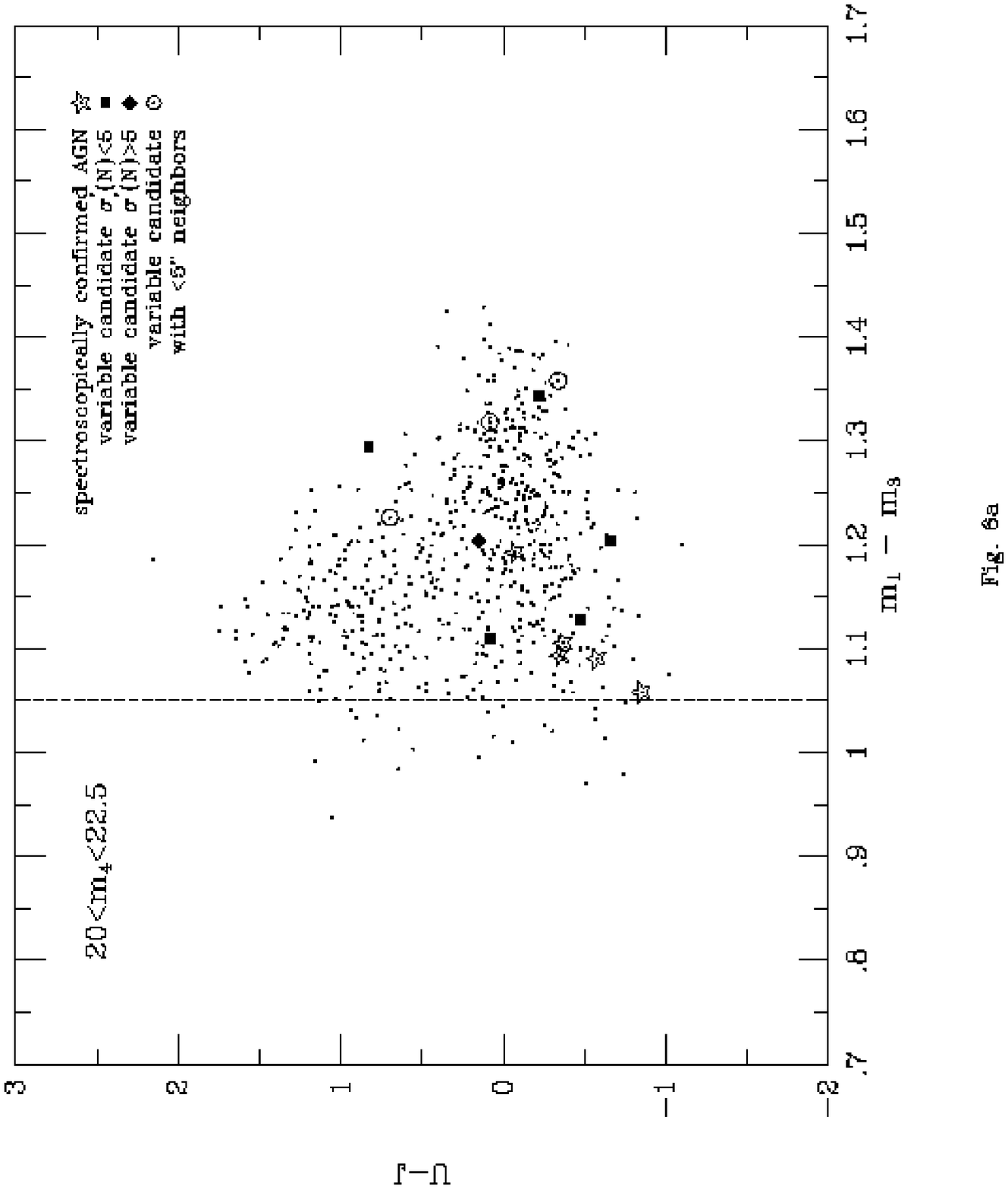}{8in}{0}{100}{100}{-315}{-75}
\end{figure}

\clearpage

\begin{figure}
\plotfiddle{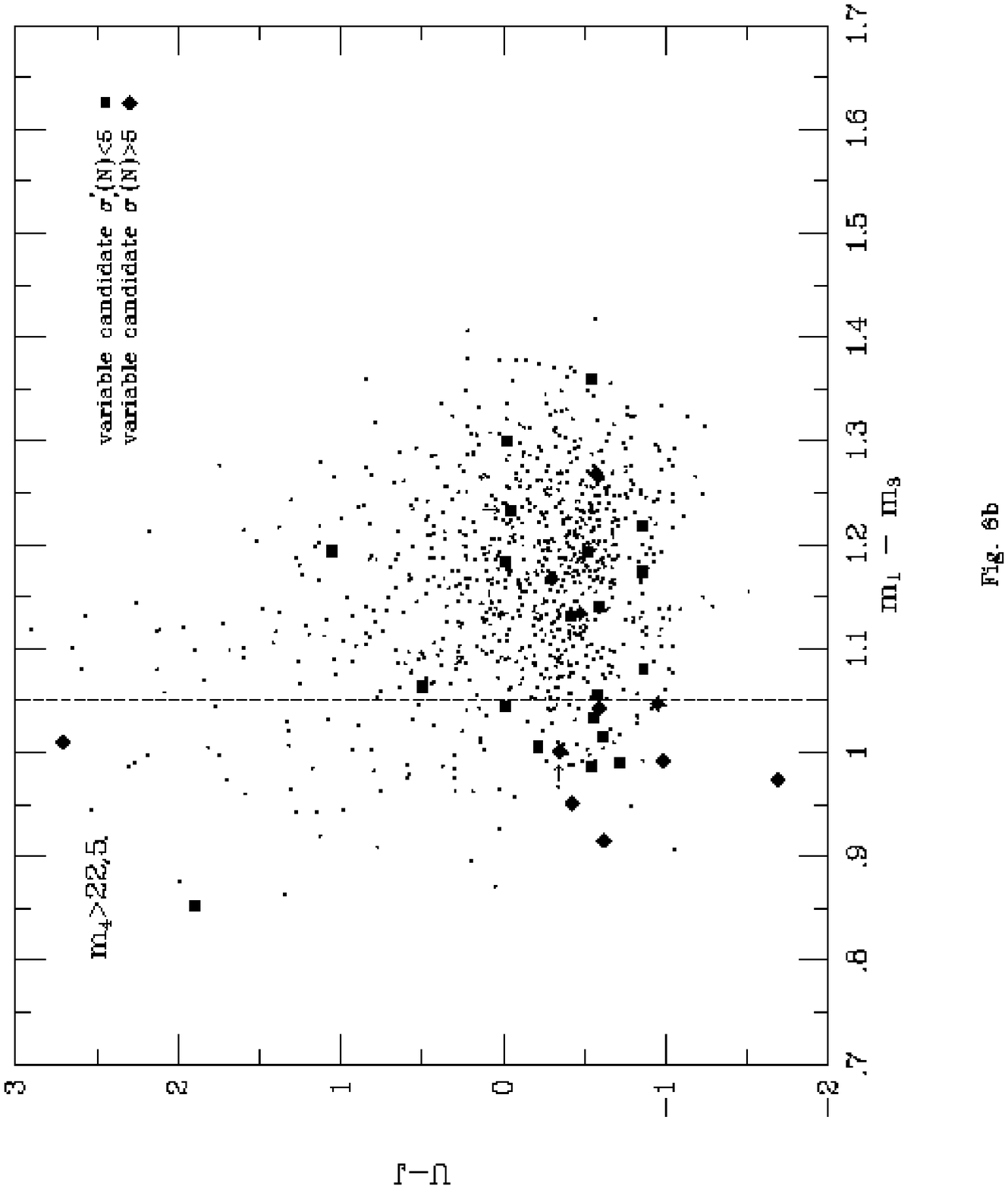}{8in}{0}{100}{100}{-315}{-75}
\end{figure}

\clearpage

\begin{figure}
\plotfiddle{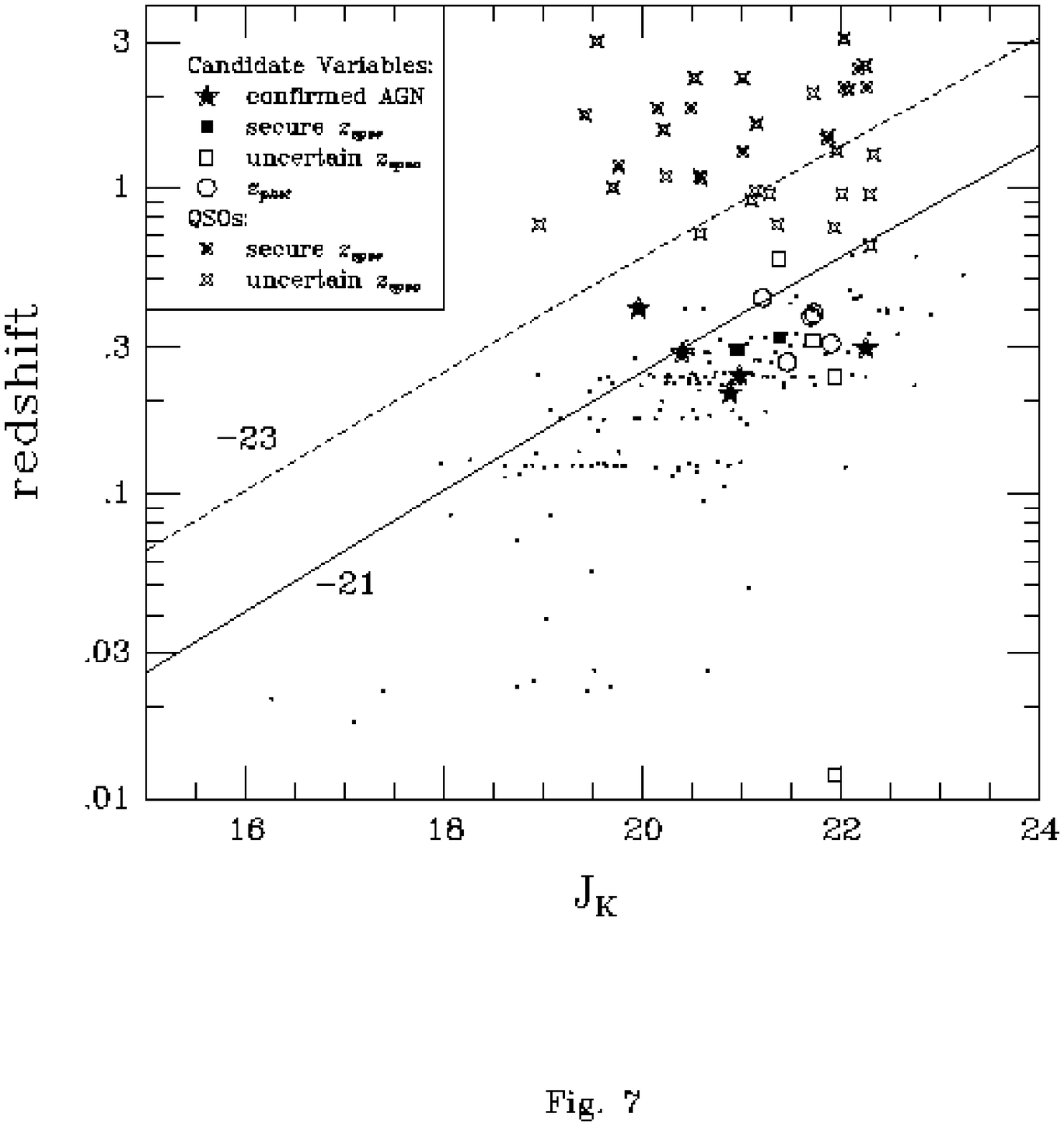}{8in}{0}{100}{100}{-305}{-130}
\end{figure}

\end{document}